\documentclass[%
 amsmath,amssymb,
 aps,prb,reprint,superscriptaddress
]{revtex4-1} 
\usepackage{braket}
\usepackage{graphicx}
\usepackage{dcolumn}
\usepackage{bm}
\usepackage{subfigure} 
\usepackage{color}

\begin{document}
\title{A Landauer formulation of photon transport in driven systems}

\author{Chiao-Hsuan Wang}
\affiliation{Joint Quantum Institute, College Park, Maryland 20742, USA}
\affiliation{University of Maryland, College Park, Maryland 20742, USA}
\affiliation{Joint Center for Quantum Information and Computer Science, College Park, Maryland 20742, USA}
\author{Jacob M. Taylor}
\affiliation{Joint Quantum Institute, College Park, Maryland 20742, USA}
\affiliation{Joint Center for Quantum Information and Computer Science, College Park, Maryland 20742, USA}
\affiliation{National Institute
of Standards and Technology, Gaithersburg, Maryland 20899, USA}


\begin{abstract}
Understanding the behavior of light in non-equilibrium scenarios underpins much of quantum optics and optical physics. While lasers provide a severe example of a non-equilibrium problem, recent interests in the near-equilibrium physics of so-called photon `gases', such as in Bose condensation of light or in attempts to make photonic quantum simulators, suggest one re-examine some near-equilibrium cases. Here we consider how a sinusoidal parametric coupling between two semi-infinite photonic transmission lines leads to the creation and flow of photons between the two lines. 
Our approach provides a photonic analog to the Landauer transport formula, and using non-equilbrium Green's functions, we can extend it to the case of an interacting region between two photonic `leads' where the sinusoid frequency plays the role of a voltage bias. Crucially, we identify both the mathematical framework and the physical regime in which photonic transport is directly analogous to electronic transport and regimes in which other behavior such as two-mode squeezing can emerge.
\end{abstract}

\maketitle
\section{Introduction}

Quantum systems have dynamics that appear to be beyond the capacity of classical computers to simulate as the size of the system increases. However, a controlled quantum simulator may enable an understanding of such systems that eludes classical description, as the emulation of one system by another can take full advantage of the underlying quantum evolution~\cite{feynman82,georgescu2014quantum}.
One promising avenue for quantum simulation uses massless bosons -- typical photons -- as the constituent particles and examines the phases of matter that can arise with the inclusion of interactions between these particles \cite{aspuru2012photonic,noh2016quantum,
heidemann2007evidence,peyronel2012quantum}. Perhaps the most dramatic possibilities arise in circuit quantum electrodynamics (QED) \cite{houck2012chip}, where the Josephson effect provides a strong microwave nonlinearity, though similar improvements are now becoming available in semiconductor, molecular, and atomic nonlinearities in small optical domain cavities \cite{Hennessy2007,kim.2013.373--377,RYDBERG}. These photonic systems are particularly interesting given our ability to control the dispersion relation of the particles, including, e.g., the creation of effective mass~\cite{kasprzak2006bose,klaers2010bose} or synthetic gauge fields \cite{Raghu2008,Hafezi2011,Fang2012} as well as the character of their interaction.
As a starting point, Bose-Einstein condensation of photons has been observed in recent experiments using cavity polaritons \cite{kasprzak2006bose,balili2007bose,deng2010exciton,tosi.2012.190--194} or with dye microcavities \cite{klaers2010bose} using these ideas.

Unfortunately, the vacuum is the typical ground state for such systems, and thus efforts for quantum simulation with light have focused on driving systems far from equilibrium to provide sufficient numbers of photons. This makes predicting the dynamics and steady state behavior an outstanding challenge~\cite{keeling.2011.131--151,nissen.2012.}.
On the other hand, electronic transport theory, pioneered in the works of Landauer \cite{landauer1957spatial,landauer1970electrical,imry1999conductance}, B\"{u}ttiker \cite{buttiker1986four}, and Imry \cite{imry1999conductance}, has successfully dealt with a different problem: What is the quantum version of Ohm's law, i.e., the relationship between chemical potential difference (voltage) and particle flux (current), for describing the motion of electrons in mesoscopic systems \cite{landauer1957spatial,landauer1970electrical,buttiker1986four,imry1999conductance,caroli1971direct,meir1992landauer,haug2008quantum}?   Of particular use have been mathematical tools such as non-equilibirum Green's function methods \cite{keldysh1965diagram,langreth1976linear,meir1992landauer,haug2008quantum}, which enable predictions for systems even at large voltage bias and with strong interactions.

In this article we consider whether a photonic version of the Landauer-type transport exists, and find that for parametrically coupled semi-infinite leads (transmission lines), a natural photonic voltage arises with an associated Ohm's law-type behavior for the photon flux. Our results rely upon the most recent of several approaches for developing a photonic equivalent to this voltage-bias \cite{wurfel1982chemical,herrmann2005light,
schmitt2014observation,ries1991chemical}, including equilibriation of light coupled to electrons flowing in a diode \cite{ries1991chemical,smestad1992luminescence,berdahl1985radiant} and, more recently, parametrically coupled photonic systems \cite{hafezi2015chemical}.
Specifically, we derive the non-equilibrium transport of light under the parametric coupling scheme using non-equilibrium Green's function (NEGF) formalism. We study the photon flux as the equivalent of a current through a parametrically driven mesoscopic region, and show that the photon flux formula can be understood in the Landauer sense, as a transport from a chemical potential imbalance from the parametric coupling, with the addition of an anomalous particle-nonconserving squeezing term. Intuitively, our result connects the photon flow between a low frequency bath and an optical bath as mediated by a mesoscopic, interacting region. Thus we provide a rigorous framework for studying such near-equilibrium photonic systems without resorting to \textit{ad hoc} tools for steady-state dynamics. Furthermore, our result predicts a quantitative link between the photon flux and the Green's function, which provides a possible testing ground for photonic quantum simulations even without particle number conservation.

\section{Photon Transport through a trivial scatterer}
\begin{figure}
\includegraphics[width=2.8in]{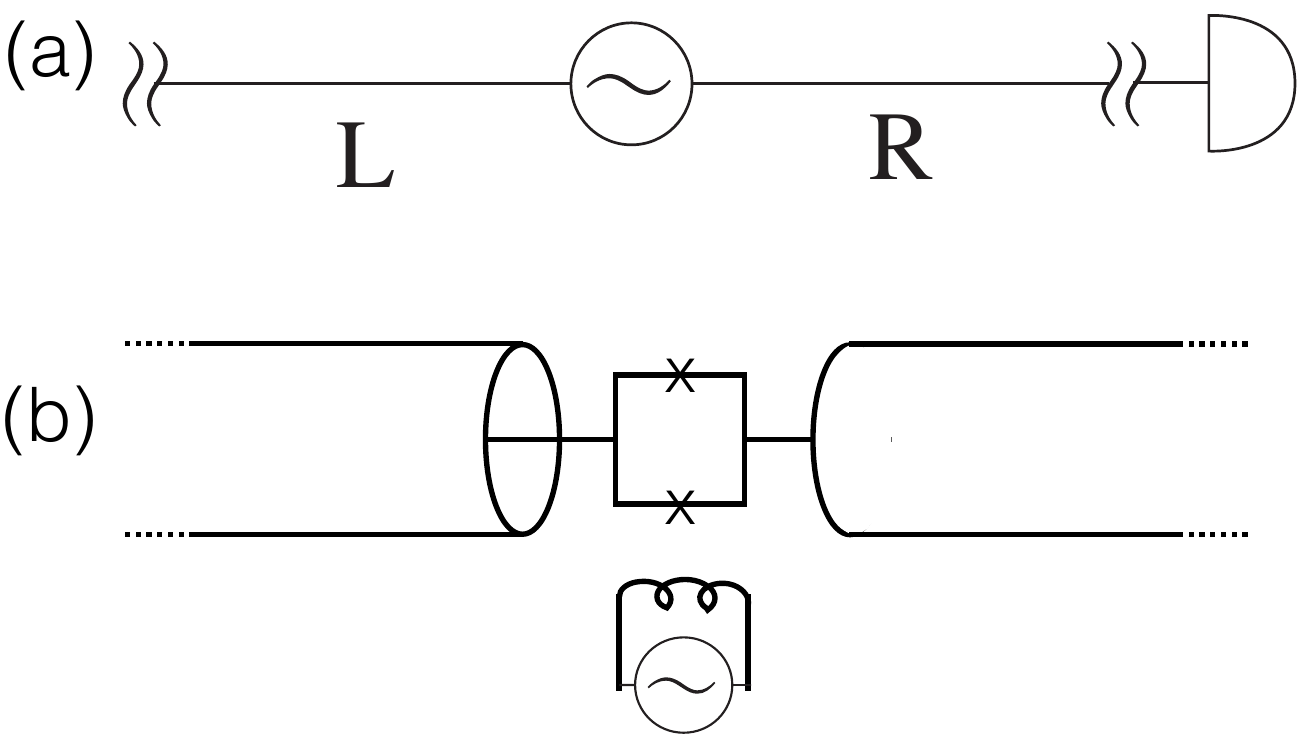}
\caption{(a) Our conceptually simplest system of two semi-infinite leads, with a time-dependent coupling between them and a photodetector connecting to the right lead. (b) A potential physical implementation with a Josephson parametric coupler, driven with a flux bias line, between two transmission lines.}
\label{interface}
\end{figure}

We start by developing our photonic analog to voltage bias. Consider a photonic (optical or microwave) system coupled to two baths: one associated with the typical decay of excitations into other modes via, e.g., imperfect mirrors, while the other is associated with a second bath coupled time-dependently with fast sinusoidal variation  of the coupling constant at angular frequency $\omega_p$. In particular, in Ref.~\cite{hafezi2015chemical}, one of us showed that a time-dependent bath coupling can lead to the equilibration of a small system best described by a grand canonical ensemble distribution, i.e., a system of photons with a chemical potential. However, in that work crucial questions -- such as what happens when coupled to two baths -- were largely detailed heuristically. Here we focus on building a formalism, analogous to the finite-bias Green's function approach for electronic transport.
In particular, we describe the two baths as semi-infinite transmission lines for our purposes, with the parametrically coupled bath being the `left' lead, and the natural bath corresponding to photon loss being the `right' lead, which could correspond to an outgoing optical signal to be measured with a photodetector (Fig.~\ref{interface}).  This is now analogous to electronic transport at finite voltage bias, where the voltage is equivalent to the chemical potential $\hbar \omega_p$.

As a toy model, and to help develop the formalism, we start with the simplest setup in which the scatterer is trivial -- a section of transmission line -- and the problem now reduces to the case with left and right semi-infinite leads coupled parametrically (see Fig.~\ref{interface}). The Hamiltonian of the the system is $H=H_L+H_R+H_T(t)$, with

\begin{align}
&H_L=\sum_{\alpha} \epsilon_{\alpha} a^{\dagger}_{\alpha} a_{\alpha}, \, H_R=\sum_{\beta} \epsilon_{\beta} b^{\dagger}_{\beta} b_{\beta}, \notag\\ &H_T(t)=\cos(\omega_p t) \sum_{\alpha,\beta} \lambda_{\alpha \beta}u_{\alpha}u_{\beta}.
\label{Hsub}
\end{align}

Here $H_L$ and $H_R$ are Hamiltonians of left and right transmission lines respectively, and $H_T(t)$ is the time-dependent tunneling coupling between the two subsystems. The summation index $\alpha$ labels the states in the left transmission line with energies $\hbar \omega_{\alpha}=\epsilon_{\alpha}$, photon annihilation operators $a_{\alpha}$, displacement operators $u_{\alpha}=\sqrt{\frac{\hbar}{2 \omega_\alpha}}\left(a^{\dagger}_{\alpha}+a_{\alpha}\right)$, and momentum operators $p_{\alpha}=i\sqrt{\frac{\hbar \omega_\alpha}{2 }}\left(a^{\dagger}_{\alpha}-a_{\alpha} \right)$, while $\beta$, $\hbar \omega_{\beta}=\epsilon_{\beta}$, $b_{\beta}$, $u_{\beta}$, and $p_{\beta}$ represent states in the right. $\lambda_{\alpha,\beta}$ are the coupling constants.

We may organize the Hamiltonian in a matrix form
\begin{align}
H=\frac{1}{2}\vec{p}^{\text{T}}\vec{p}+\frac{1}{2}\vec{u}^{\text{T}}\mathbf{K} \vec{u}
\label{Hmatrix}
\end{align}
by introducing displacement and momentum vectors
\begin{align}
\vec{u}=\begin{pmatrix} \vec{u}_L\\ \vec{u}_R \end{pmatrix}, \vec{p}=\begin{pmatrix} \vec{p}_L\\ \vec{p}_R \end{pmatrix},
\label{up}
\end{align}
with elements $(u_{L(R)})_{\alpha(\beta)}\equiv u_{\alpha(\beta)}$ and $(p_{L(R)})_{\alpha(\beta)}\equiv p_{\alpha(\beta)}$.
$\vec{u}$ and $\vec{p}$ follow the equal time commutation relation
\begin{align}
\left[ \vec{u}(t),\vec{p}^{\text{T}}(t) \right]=i\hbar \mathbf{I}.
\label{commutation}
\end{align}
Here $\mathbf{I}$ is the identity matrix, and T denotes the matrix transpose. $\mathbf{K}$ is a symmetric spring's constant matrix and can be further separated into diagonal and off-diagonal parts $\mathbf{K}=\mathbf{D}+\mathbf{V}(t)=\mathbf{D}+\mathbf{V}\cos(\omega_pt)$, where 
\begin{align}
\mathbf{D}=\begin{pmatrix} \mathbf{D}^L & 0\\ 0 & \mathbf{D}^R \end{pmatrix}, \mathbf{V}(t)=\begin{pmatrix} 0 &\mathbf{V}^{LR}(t)\\ \mathbf{V}^{RL}(t) & 0 \end{pmatrix}.
\label{DV}
\end{align}
Here $D^L_{\alpha \alpha '}=\omega^2_{\alpha} \delta_{\alpha \alpha '}$, $D^R_{\beta \beta '}=\omega^2_{\beta} \delta_{\beta \beta '}$, and $V^{LR}_{\alpha \beta}(t)=V^{RL}_{\beta \alpha}(t)=\cos(\omega_p t) \lambda_{\alpha \beta}.$

Assume the two subsystems were initially decoupled and in their own thermal equilibrium, and the parametric coupling is adiabatically turned on at $t = - \infty$ and turned off at $t = \infty$. Our goal is to find the photonic current transported between the two ends and express it in a Landauer-like formula in order to predict the current based on an effective chemical potential difference analogous to a voltage bias.

The current on the right at some later time $t$ is defined as the temporal change of the total number of photons in the right transmission line $N_R=\sum_{\beta} b_{\beta}^{\dagger} b_{\beta}$, which corresponds to an expected photodetector signal. We have
\begin{align}
J_R(t) \equiv \left\langle \dot{N}_R(t) \right\rangle =\left\langle \frac{d \sum_{\beta} b_{\beta}^{\dagger} b_{\beta} (t)}{dt}\right\rangle.
\label{JR}
\end{align}
The angular bracket denotes ensemble average over the initial equilibrium density of states, while the operators are in the Heisenberg picture.
According to the Heisenberg equation of motion,
\begin{align}
\dot{N}_R(t)=-\frac{1}{\hbar} \left. \left( \frac{\partial}{\partial t^{'}}\left[ \vec{u}_R^{\text{T}}(t {'}) \tilde{\mathbf{V}}^{RL}(t) \vec{u}_L(t)\right] \right) \right|_{t {'}=t},
\label{Ndot}
\end{align}
where $\tilde{V}^{RL}_{ \beta \alpha} (t) \equiv V^{RL}_{ \beta \alpha} (t) /\omega_{\beta}=\cos(\omega_p t)\lambda_{\alpha \beta}/\omega_{\beta}.$

One can connect the current expression with Keldysh Green's functions \cite{keldysh1965diagram} by introducing the non-equilibrium lesser Green's function defined as
\begin{align}
\mathbf{G}^{<}(t,t {'})\equiv -\frac{i}{\hbar} \left\langle \vec{u}(t  {'}) \vec{u}^{\text{T}}(t)\right\rangle ^{\text{T}},
\label{Gless}
\end{align}
which can also be split into four blocks associated with left and right transmission lines. We can write the current using the lesser Green's function as
\begin{align}
J_R(t)=-i \left. \left( \frac{\partial}{\partial t^{'}}\text{Tr}\left[ \mathbf{G}^{<}_{LR}(t,t{'}) \tilde{\mathbf{V}}^{RL}(t)\right] \right) \right|_{t {'}=t}.
\label{JLinG}
\end{align}
The trace here means tracing over photon states $\alpha$.

We now follow the standard Keldysh formalism (NEGF formalism) \citep{keldysh1965diagram} to study the transport formula \cite{caroli1971direct,meir1992landauer,haug2008quantum,wang2014nonequilibrium}. Since we define our Green's functions on displacement operators $u$ instead of photon creation operators $a^{\dagger}$, our problem structurally resembles more the thermal transport cases \citep{wang2014nonequilibrium} than electronic ones. We remind the reader here that since the parametric coupling varies with time and allows pair production and annihilation mechanisms, many identities and tricks in previous works involving steady state or particle-conserving assumptions cannnot be applied here.

The equation of motion of the contour ordered Green's function defined on the Keldysh contour $C$ follows
\begin{align}
\frac{\partial^2}{\partial \tau^2} \mathbf{G}^c(\tau, \tau {'})+\mathbf{K} \mathbf{G}^c(\tau, \tau {'})=-\delta(\tau, \tau {'}) \mathbf{I},
\label{GEOM}
\end{align}
while the noninteracting equilibrium Green's function $\mathbf{g}^c(\tau, \tau {'})$ follows the equation of motion
\begin{align}
\frac{\partial^2}{\partial \tau^2} \mathbf{g}^c(\tau, \tau {'})+\mathbf{D} \mathbf{g}^c(\tau, \tau {'})=-\delta(\tau, \tau {'}) \mathbf{I}.
\label{gEOM}
\end{align}
One can easily verify that $\mathbf{G}^c(\tau, \tau {'})$ follows the Dyson equation
\begin{align}
\mathbf{G}^c(\tau, \tau {'})=\mathbf{g}^c(\tau, \tau {'})+\int_{C} d \tau {''}\mathbf{g}^c(\tau, \tau {''}) \mathbf{V}(\tau {''}) \mathbf{G}^c(\tau {''}, \tau {'}).
\label{Dyson}
\end{align}

Using the Langreth theorem of analytic continuation \cite{langreth1976linear}, the lesser Green's function can be expressed as an integral form on the real axis
\begin{align}
\mathbf{G}^<_{LR}(t, t {'}) & \approx  \int_{-\infty}^{\infty} dt_1  \left\{ \mathbf{g}^r_L(t, t_1) \mathbf{V}^{LR}(t_1) \mathbf{g}^<_R(t_1, t {'})  \right. \notag\\
& + \left. \mathbf{g}^<_L(t, t_1) \mathbf{V}^{LR}(t_1) \mathbf{g}^a_R(t_1, t {'})\right\} +\mathcal{O}(\lambda^2)
.
\label{Glessfin}
\end{align}
Here the $r$ and $a$ superscripts stand for retarded and advanced Green's functions, and we treat $\lambda$ as a perturbation.
The equilibrium Green's functions used in the $\mathbf{G}^<_{LR}(t, t {'})$ expression are given by
\begin{align}
 &(g^r_L)_{\alpha}(t, t_1)=\frac{-i}{2 \omega_{\alpha}}\theta(t-t_1) \left( e^{-i \omega_{\alpha} (t-t_1)}-e^{i \omega_{\alpha} (t-t_1)}\right),\notag\\
 &(g^<_R)_{\beta}(t_1, t{'})=\frac{-i}{2 \omega_{\beta}}\left\{ n_R(\epsilon_{\beta}) e^{-i \omega_{\beta}(t_1-t{'})}+[1+n_R(\epsilon_{\beta})] e^{i \omega_{\beta}(t_1-t{'})}\right\},\notag\\
  &(g^<_L)_{\alpha}(t, t_1)=\frac{-i}{2 \omega_{\alpha}}\left\{ n_L(\epsilon_{\alpha}) e^{-i \omega_{\alpha}(t-t_1)}+[1+n_L(\epsilon_{\alpha})] e^{i \omega_{\alpha}(t-t_1)}\right\},\notag\\
 &(g^a_R)_{\beta}(t_1, t {'})=\frac{-i}{2 \omega_{\beta}}\theta(t{'}-t_1) \left( e^{i \omega_{\beta} (t_1-t{'})}-e^{-i \omega_{\beta} (t_1-t{'})}\right).\notag\\
\label{g}
\end{align}
Here $n_{L(R)}(\epsilon_{\alpha (\beta)})=(e^{(\epsilon_{\alpha (\beta)}-\mu_{L(R)})/k_BT}-1)^{-1}$ are the bosonic occupation number in left and right transmission lines. The chemical potentials $\mu_L=\mu_R=0$ for photons, $k_B$ is the Boltzmann constant, and $T$ is the initial temperature of the system.

By inserting the expression for $\mathbf{G}^<_{LR}(t, t {'})$, the current is now
\begin{widetext}
\begin{align}
J_R(t)=&-i \left( \left.\frac{\partial}{\partial t^{'}}\text{Tr}\left[ \int_{-\infty}^{\infty} dt_1 \left\{ \mathbf{g}^r_L(t, t_1) \mathbf{V}^{LR}(t_1) \mathbf{g}^<_R(t_1, t {'})+ \mathbf{g}^<_L(t, t_1) \mathbf{V}^{LR}(t_1) \mathbf{g}^a_R(t_1, t {'})\right \} \tilde{\mathbf{V}}^{RL}(t)\right] \right) \right|_{t {'}=t} \notag\\
=&-i \left( \left. \frac{\partial}{\partial t^{'}}\sum_{\alpha,\beta} \int_{-\infty}^{\infty} dt_1 \frac{\lambda_{\alpha \beta}^2}{\omega_{\beta}} \cos(\omega_p t_1)\cos(\omega_p t)\left\{ (g^r_L)_{\alpha}(t, t_1) (g^<_R)_{\beta}(t_1, t {'})+ (g^<_L)_{\alpha}(t, t_1) (g^a_R)_{\beta}(t_1, t {'})\right \}\right) \right|_{t {'}=t} \notag\\
=&\sum_{\alpha,\beta} \frac{\lambda_{\alpha \beta}^2}{4\omega_{\alpha}\omega_{\beta}} \int_{0}^{\infty} d \tau \left( \left\{\cos(\omega_p \tau)+\cos[\omega_p (2t-\tau)]\right\}\cos[(\omega_{\beta}-\omega_{\alpha})\tau]\left[n_L(\epsilon_{\alpha})-n_R(\epsilon_{\beta})\right] \right. \notag\\
&+ \left. \left\{\cos(\omega_p \tau)+\cos[\omega_p (2t-\tau)]\right\}\cos[(\omega_{\beta}+\omega_{\alpha})\tau]\left[ n_L(\epsilon_{\alpha})+n_R(\epsilon_{\beta})+1 \right] \right).
\label{JR1}
\end{align}
\end{widetext}

In the last equality we have used the identity $\cos(\omega_p t_1)\cos(\omega_p t)=\left\{\cos(\omega_p \tau)+\cos[\omega_p (2t-\tau)]\right\}/2$, and changed the integral variable to $\tau=t-t_1.$ The only explicit $t$ dependence arises in the $\cos[\omega_p (2t-\tau)]$ factor. Averaging over one pump cycle $2 \pi/ \omega_p$ takes this factor to zero. We thus neglect those terms with $\cos[\omega_p (2t-\tau)]$ in the spirit of the rotating wave approximation.

We assume the coupling constant only depends on the mode energy, $\lambda_{\alpha \beta}=\lambda(\epsilon_{\alpha},\epsilon_\beta)$, and take the continuum limit of energy so that $\sum_{\alpha,\beta}=\int_{0}^{\infty} d \epsilon_{\alpha} \rho_L(\epsilon_{\alpha})\int_{0}^{\infty} d \epsilon_{\beta} \rho_R(\epsilon_{\beta})$. Here $\rho_L$ and $\rho_R$ are the energy density of states in the left and right transmission lines. Note that $\int_0^{\infty} \cos[(\omega-\omega_1)\tau] d\tau=\pi \delta(\omega-\omega_1)$.
We now arrive at a current formula with three terms:
\begin{align}
\bar{J}_R=&\int_{\hbar \omega_p}^{\infty} d \epsilon T (\epsilon,\epsilon-\hbar \omega_p) \left[  n_L(\epsilon)-n_R(\epsilon-\hbar \omega_p)\right] \notag\\
+& \int_{\hbar \omega_p}^{\infty} d \epsilon  T(\epsilon-\hbar \omega_p,\epsilon) \left[ n_L(\epsilon-\hbar \omega_p)-n_R(\epsilon) \right] \notag\\
+& \int_0^{\hbar \omega_p} d \epsilon T(\epsilon,\hbar \omega_p-\epsilon) \left[ n_L(\epsilon)+n_R(\hbar \omega_p-\epsilon)+1 \right] 
\label{JRfinal}
\end{align}

Here $\bar{J}_R$ represents the time-averaged current under the rotating wave approximation, and $T(\epsilon_1,\epsilon_2)$ is the transmission function defined as $T(\epsilon_1,\epsilon_2)=\frac{\hbar^3 \pi}{8 \epsilon_1 \epsilon_2} \lambda^2(\epsilon_1,\epsilon_2)\rho_L(\epsilon_1)\rho_R(\epsilon_2).$ $\bar{J}_L$ can be calculated with similar formulations.

Note that the $1/\epsilon$ factor in the transmission function and the bosonic occupation numbers $n_{L(R)}(\epsilon)$ go to infinity as the photon energy approaches zero. One can ensure the convergence of our model by the choice of a three-dimensional reservoir on the low-frequency side. The presence of the nonlinear interaction in the case of an interacting mesoscopic region can regulate the problem as well \cite{hafezi2015chemical}. The power from the pump that generates the parametric coupling should be finite, and as the IR divergence is approached for lower dimensional systems, an appropriate inclusion of pump depletion will be necessary to develop a complete understanding of the problem.

The first line of Eq. (\ref{JRfinal}) can be interpreted as a Landauer-like transport with an effective chemical potential $\hbar \omega_p$ on the right transmission line, and the second line represents a Landauer-like transport with an effective chemical potential $\hbar \omega_p$ on the left. The third line is a particle-nonconserving term due to pair creation and annihilation mechanisms allowed by the oscillating $u$-$u$ type coupling. Two-mode squeezed states of light \cite{milburn1984multimode} are generated through this mechanism with the photon pairs entangled. One will expect a thermal state when tracing over the output modes on one side of such photon pairs.

Note that the current formula is consistent with Fermi's golden rule: The parametric coupling $\cos(\omega_p t)$ only allows transition with $E_f-E_i=\Delta E=\pm \hbar \omega_p$, where $E_f$ and $E_i$ are the energies of the final and initial states. For $\omega_p=0$, the current equation reduces to the usual Landauer form proportional to $n_L(\epsilon)-n_R(\epsilon),$ which is essentially zero when the two transmission lines are at the same temperature.

The particle-nonconserving nature of the problem is manifested by identifying the anomalous current $\bar{J_A}\equiv (\bar{J}_R+\bar{J}_L)/2=\int_0^{\hbar \omega_p} d \epsilon T(\epsilon,\hbar \omega_p-\epsilon) \allowbreak \left[ n_L(\epsilon)+n_R(\hbar \omega_p-\epsilon)+1 \right]$, which is only zero when $T(\epsilon,\hbar \omega_p-\epsilon)=0$ throughout the range, as is the case in Fig. \ref{center}(b). This term can also be understood in Fermi's golden rule point of view, considering the harmonic perturbation $H_T(t)$. According to Fermi's golden rule, the pair creation (annihilation) rates $R_c$ ($R_a$) are:
\begin{align}
&R_c= \int_0^{\hbar \omega_p} d \epsilon T(\epsilon,\hbar \omega_p-\epsilon) \left[ n_L(\epsilon)+1\right] \left[n_R(\hbar \omega_p-\epsilon)+1 \right], \notag\\
&R_a= \int_0^{\hbar \omega_p} d \epsilon T(\epsilon,\hbar \omega_p-\epsilon) n_L(\epsilon) n_R(\hbar \omega_p-\epsilon).
\label{RcRa}
\end{align}
The net creation rate is thus $R_c-R_a=\bar{J}_A.$

One can find the non-equilibrium transport part of the current by subtracting the anomalous squeezing (particle-nonconversing) term $\bar{J}_A$, and we are left with the normal current $\bar{J}_N\equiv (\bar{J}_R-\bar{J}_L)/2$, the first two lines of Eq. (\ref{JRfinal}). We note here that the asymmetry between right and left is necessary for the transport to occur; the first two lines of Eq. (\ref{JRfinal}) will cancel each other otherwise.

To focus on the transport mechanism only, we consider an energy gap on the right transmission line [see Fig.~\ref{center}(b)] such that $\forall \alpha, \beta, \epsilon_{\beta}>\hbar \omega_p,\epsilon_{\beta}>\epsilon_{\alpha}$. This gap setup prevents the pair creation and annihilation mechanisms, leaves us with a conserved current and permits a direct photonic analog to electronic transport. The system follows the transport formula  $\bar{J}_R=\int_{\epsilon_{\beta,\rm min}}^{\epsilon_{\alpha,\rm max}+\hbar \omega_p} d \epsilon T (\epsilon-\hbar \omega_p,\epsilon) \left[  n_L(\epsilon-\hbar \omega_p)-n_R(\epsilon)\right]$, which is equivalent to a non-equilibirum transport current under a chemical potential imbalance $\mu_L=\hbar \omega_p, \mu_R=0.$

One can see the resemblance between our gapped transport equation and the I-V (current-voltage) characteristic of an ideal light emitting diode (LED) \cite{ries1991chemical} by relating the chemical potential $\hbar \omega_p$ to $qV$, and the gap energy $\epsilon_{\beta,\rm min}$ to the photon energy threshold $e_g$, and working under the region $\epsilon_{\alpha,\rm max}+\hbar\omega_p \gg  k_B T$. However, we cannot yet make a direct connection mathematically with the somewhat different problem of electron transport through a diode combined with emission of photon into an interacting region.

\section{Photon Transport through a mesoscopic central region}

\begin{figure}
\includegraphics[width=3.0in]{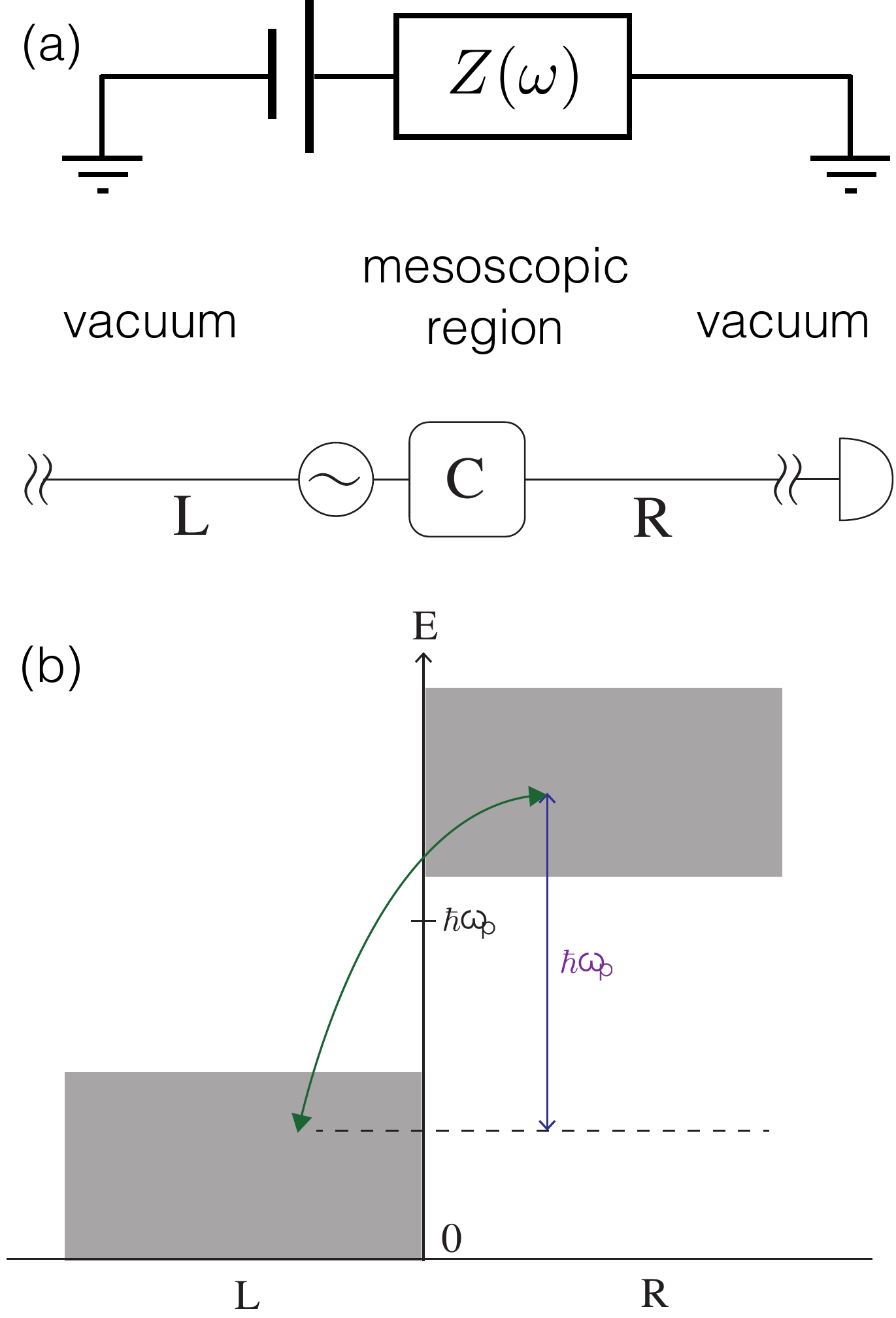}
\caption{(a) The generic mesoscopic scenario, with two semi-infinite leads coupled parametrically to an intermediate mesoscopic region $C$ provides a photonic equivalent to a voltage bias and a (nonlinear) impedance provided by the region $C$. (b) A schematic diagram of the parametric coupling mechanism, in which a low energy photonic mode on the left side is up-converted via the pump photon with energy $\hbar \omega_p$ to a higher energy mode on the right. The case where $L$ and $R$ leads have a high (left) and low (right) frequency cutoff is shown, to prevent anomalous squeezing terms.}
\label{center}
\end{figure}

Now we consider a more generic case with a center mesoscopic region placed between the transmission lines, with parametric coupling between the center region and the left transmission line (see Fig.~\ref{center}). We replace the time-dependent barrier $H_T(t)$ with $H_C+H_{CL}(t)+H_{CR}$, and the Hamiltonian becomes

\begin{align}
H=H_L+H_C+H_R+H_{CL}(t)+H_{CR},
\label{CH}
\end{align} 
\begin{align}
& H_L=\sum_{\alpha} \epsilon_{\alpha} a^{\dagger}_{\alpha} a_{\alpha}, \, H_R=\sum_{\beta} \epsilon_{\beta} b^{\dagger}_{\beta} b_{\beta},\notag \\
&H_C= \sum_{\gamma,\gamma {'}} t_{\gamma \gamma {'}} c^{\dagger}_{\gamma} c_{\gamma '}+H_{\rm int}, \, \notag \\
&H_{CL}(t)=\cos(\omega_p t) \sum_{\alpha,\gamma} \lambda_{\alpha \gamma}u_{\alpha}u_{\gamma}, \notag\\
&H_{CR}=\sum_{\beta,\gamma} \breve{\lambda}_{\beta \gamma}u_{\beta}u_{\gamma}.
\label{CHsub}
\end{align}
The summation index $\gamma$ labels the states in the center with energies $\hbar \omega_{\gamma}=\epsilon_{\gamma}$ and photon annihilation operators $c_{\gamma}$. Note that we place all time-dependence in the center to left coupling $H_{CL}(t)$. The central region can contain some nonlinear interacting term $H_{\rm int}$ as well as non-trivial single-particle potential effects from $t_{\gamma\gamma'}$.
We again assume the subsystems were initially in their own thermal equilibrium before the parametric coupling  adiabatically turned on at $t = - \infty$.

The current on the right can be expressed as
\begin{align}
J_R(t)=-i \left( \left. \frac{\partial}{\partial t^{'}}\text{Tr}\left[ \mathbf{G}^{<}_{CR}(t,t{'}) \tilde{\mathbf{V}}^{RC}\right]\right) \right|_{t {'}=t}.
\label{CJRinG}
\end{align}
Here $\tilde{V}^{RC}_{\beta \gamma}=V^{RC}_{\beta \gamma}/ \omega_{\beta}=\breve{\lambda}_{\beta \gamma} / \omega_{\beta}.$

Accordingly to the Dyson equation, $\mathbf{G}^c_{CR}(\tau, \tau {'})=\int_{C} \mathbf{G}_{CC}^c(\tau, \tau {''}) \mathbf{V}^{CR}\mathbf{g}^c_R(\tau {''}, \tau {'}).$ Using the Langreth theorem of analytic continuation and defining $\mathbf{\tilde{\Sigma}}_R(t_1,t{'})=\mathbf{V}^{CR}\mathbf{g}_R(t_1,t{'}) \mathbf{\tilde{V}}^{RC},$ we have
\begin{align}
J_R(t)=-i \int_{-\infty}^{\infty} dt_1 \Big( {\frac{\partial}{\partial t '}}\text{Tr} & \left[  \mathbf{G}^r_{CC}(t, t_1)  \mathbf{\tilde{\Sigma}}^<_R(t_1, t {'})+ \right. \notag\\
& \left. \left. \left. \mathbf{G}^<_{CC}(t, t_1) \mathbf{\tilde{\Sigma}}^a_R(t_1, t {'})\right] \right) \right|_{t {'}=t}.
\label{CJRgeneral}
\end{align}
This is our generic current expression analogous to the Meir-Wingreen equation in the electronic transport theory \cite{meir1992landauer}.

One can reformulate the current through Fourier transform
\begin{align}
J_R(t)=\int_{-\infty}^{\infty} \frac{\omega d \omega}{2 \pi}  \text{Tr}\left[  \mathbf{G}^r_{CC}(t, \omega)  \mathbf{\tilde{\Sigma}}^<_R(\omega)+ \mathbf{G}^<_{CC}(t, \omega) \mathbf{\tilde{\Sigma}}^a_R(\omega)\right].
\label{CJRFT}
\end{align}
Here $\mathbf{G}_{CC}(t, t{'}) \equiv \int_{-\infty}^{\infty} \frac{d \omega}{2 \pi} e^{-i \omega (t-t{'})} \mathbf{G}_{CC}(t,\omega )$, $\mathbf{G}_{CC} (t, \omega ) \equiv \int_{-\infty}^{\infty} d t {'} e^{ i \omega (t-t{'})} \mathbf{G}_{CC} (t, t{'})$.
We note here that due to the time-dependent coupling, the center Green's function is not in a steady state, and its Fourier transform function therefore depends on the initial (or ending) time index.

We now examine the simplified case of non-interacting center, and the main result of this work follows through. Non-interacting mesoscopic transport theory provides interesting phenomena such as weak localization, ballistic-to-diffusive transitions, and weak-antilocalization \cite{lee1985disordered,imry2002introduction,akkermans2007mesoscopic}. For $H_{\rm int}=0$, the center Green's function follows the Dyson equation
\begin{align}
\mathbf{G}^c_{CC}(\tau, \tau {'})=\mathbf{g}^c_C(\tau, \tau {'})+\int_{C} \mathbf{g}_{C}^c(\tau, \tau_1) \mathbf{\Sigma}^c_{\rm tot}(\tau_1,\tau_2) \mathbf{G}_{CC}^c(\tau_2, \tau {'}).
\label{CDysonBallistic}
\end{align}
Here $\mathbf{\Sigma}^c_{\rm tot}(\tau_1,\tau_2)$ is the total self-energy of the center,
\begin{align}
&\mathbf{\Sigma}^c_{\rm tot}(\tau_1,\tau_2)=\mathbf{\Sigma}^c_L(\tau_1,\tau_2)+\mathbf{\Sigma}^c_R(\tau_1,\tau_2), \notag\\
 &\mathbf{\Sigma}^c_L(\tau_1,\tau_2)\equiv \mathbf{V}^{CL}\cos(\omega_p \tau_1)\mathbf{g}^c_L(\tau_1,\tau_2)\cos(\omega_p \tau_2)\mathbf{V}^{LC},\notag\\
   &\mathbf{\Sigma}^c_R(\tau_1,\tau_2)\equiv \mathbf{V}^{CR}\mathbf{g}^c_R(\tau_1,\tau_2)\mathbf{V}^{RC}.
\label{CSigma}
\end{align} 
Note that for the case of interacting center, there will be additional contribution to the self-energy depending on the details of $H_{\rm int}$.

Specifically, the center greater and lesser Green's function follows
\begin{align}
\mathbf{G}^{\stackrel{>}{<}}_{CC}(t, t {'})=\int dt_1 dt_2 \mathbf{G}^r_{CC}(t, t_1) \mathbf{\Sigma}^{\stackrel{>}{<}}_{\rm tot}(t_1,t_2) \mathbf{G}^a_{CC}(t_2, t {'}),
\end{align}
\label{CG><}
which allows us to further simplify the current expression.

The center Green's function under parametric coupling can be expanded with the harmonics of the coupling frequency $\omega_p$. Specifically, $\mathbf{G}_{CC} (t, \omega )=\sum_n\mathbf{G}_{CC,n} (\omega) e^{2 n i \omega_p t}, n \in \mathbb{Z}$. Under the rotating wave approximation, we neglect fast oscillating terms with $n \neq 0$ and keep only the $n=0$ steady part of the current. The time averaged current is now
\begin{align}
\bar{J}_R=\int_{-\infty}^{\infty} \frac{d \omega}{2 \pi} \omega \text{Tr}\left[  \mathbf{G}^r_{CC,0}(\omega)  \mathbf{\tilde{\Sigma}}^<_R(\omega)+ \mathbf{G}^<_{CC,0}(\omega) \mathbf{\tilde{\Sigma}}^a_R(\omega)\right].
\label{CJRRWA}
\end{align}

Since the current is real, $\bar{J}_R=(\bar{J}_R+\bar{J}^*_R)/2$. Using the general identity $G^>-G^<=G^r-G^a$ and identities for steady state Green's functions in the frequency domain $[G^r(\omega)]^{\dagger}=G^a(\omega)$, $[G^<(\omega)]^{\dagger}=-G^<(\omega)$,  
\begin{align}
\bar{J}_R=\int_{-\infty}^{\infty} \frac{d \omega}{4 \pi} \omega \text{Tr} & \left\{  \left[\mathbf{G}^>_{CC,0}(\omega)-\mathbf{G}^<_{CC,0}(\omega)\right]  \mathbf{\tilde{\Sigma}}^<_R(\omega) \right. \notag\\
& \left.-\mathbf{G}^<_{CC,0}(\omega) \left[\mathbf{\tilde{\Sigma}}^r_R(\omega)-\mathbf{\tilde{\Sigma}}^a_R(\omega)\right]\right\}.
\label{CJRRWA2}
\end{align}

Expanding the non-interacting center greater and lesser Green's functions to the leading order term yields: 
\begin{align}
\mathbf{G}^{\stackrel{>}{<}}_{CC}(t, t {'})  \approx& \int dt_1 dt_2 \mathbf{g}^r_{CC}(t, t_1) \mathbf{\Sigma}^{\stackrel{>}{<}}_{\rm tot}(t_1,t_2) \mathbf{g}^a_{CC}(t_2, t {'})\notag\\
&+\mathcal{O}(\lambda^3), \notag\\ 
\mathbf{G}^{\stackrel{>}{<}}_{CC,0}(\omega) \approx &\mathbf{g}^r_{C}(\omega) \mathbf{\Sigma}^{\stackrel{>}{<}}_{\rm tot,0}(\omega) \mathbf{g}^a_{C}(\omega)  \notag\\
=& \mathbf{g}^r_{C}(\omega) \left( \mathbf{V}^{CR} \mathbf{g}^{\stackrel{>}{<}}_R(\omega)\mathbf{V}^{RC}\right.  \notag\\
&+\frac{1}{4}\mathbf{V}^{CL} \mathbf{g}^{\stackrel{>}{<}}_L(\omega+\omega_p)\mathbf{V}^{LC}\notag\\
&+\left.\frac{1}{4}\mathbf{V}^{CL} \mathbf{g}^{\stackrel{>}{<}}_L(\omega-\omega_p)\mathbf{V}^{LC} \right) \mathbf{g}^a_{C}(\omega).
\label{CG><p}
\end{align}

By inserting the equilibrium Green's function for left and right transmission lines $g_L$ and $g_R$, we arrive at a formula similar to the trivial scatterer problem
\begin{align}
\bar{J}_R=&\int_{\hbar \omega_p}^{\infty} d \epsilon T_C (\epsilon,\epsilon-\hbar \omega_p) \left[  n_L(\epsilon)-n_R(\epsilon-\hbar \omega_p)\right] \notag\\
+& \int_{\hbar \omega_p}^{\infty} d \epsilon  T_C(\epsilon-\hbar \omega_p,\epsilon) \left[ n_L(\epsilon-\hbar \omega_p)-n_R(\epsilon) \right] \notag\\
+& \int_0^{\hbar \omega_p} d \epsilon T_C(\epsilon,\hbar \omega_p-\epsilon) \left[ n_L(\epsilon)+n_R(\hbar \omega_p-\epsilon)+1 \right] .
\label{CJRfinal}
\end{align}
where the center transmission function is
\begin{align}
&T_C(\epsilon_{\alpha},\epsilon_{\beta})=\frac{\pi \hbar^3}{8}\text{Tr}\left[ \mathbf{g}_c^r(\epsilon_{\beta})\mathbf{\Lambda}_R(\epsilon_{\beta})\mathbf{g}_c^a(\epsilon_{\beta})\mathbf{\Lambda}_L(\epsilon_{\alpha})\right], \notag\\
&[\Lambda_L(\epsilon_{\alpha})]_{\gamma_1,\gamma_2}=\rho_L(\epsilon_{\alpha})\lambda_{\gamma_1}(\epsilon_{\alpha})\lambda_{\gamma_2}(\epsilon_{\alpha})/\epsilon_{\alpha},\notag \\ &[\Lambda_R(\epsilon_{\beta})]_{\gamma_1,\gamma_2}=\rho_R(\epsilon_{\beta})\breve{\lambda}_{\gamma_1}(\epsilon_{\beta})\breve{\lambda}_{\gamma_2}(\epsilon_{\beta})/\epsilon_{\beta}.
\end{align}
\label{CTC}

We again have the first two lines of Eq. (\ref{CJRfinal}) as the Landauer-like transport terms, and the last line is the particle-nonconserving part due to the oscillating $u$-$u$ type coupling.
The system will undergo non-equilibirum transport with an effective chemical potential imbalance $\hbar \omega_p$ under specific gap setups, and the current expression resembles the I-V characteristic of an ideal light-emitting diode.

\section{Conclusions}
We have derived the photonic flux between different baths parametrically-coupled to an intermediate system and found a Landauer-like transport formula for non-interacting centers. However, we also have another regime, with a particle-nonconserving term, which we can interpret as a two-mode squeezing output. The consequences of this latter regime for observation and even application remain to be explored, and are beyond the scope of the present work. We have also shown a potential extension of these techniques at the formal level to the interacting case, but suggest that applying these results, e.g., to photon-blockaded systems to see the non-classical light output would be an intriguing next step.

\begin{acknowledgments}
Funding is provided by NSF Physics Frontier Center at the JQI, and from the Princeton Center for Complex Materials, a MRSEC supported by NSF Grant No. DMR 1420541.
\end{acknowledgments}



\bibliography{transportbibV2}

\begin{thebibliography}{41}%
\makeatletter
\providecommand \@ifxundefined [1]{%
 \@ifx{#1\undefined}
}%
\providecommand \@ifnum [1]{%
 \ifnum #1\expandafter \@firstoftwo
 \else \expandafter \@secondoftwo
 \fi
}%
\providecommand \@ifx [1]{%
 \ifx #1\expandafter \@firstoftwo
 \else \expandafter \@secondoftwo
 \fi
}%
\providecommand \natexlab [1]{#1}%
\providecommand \enquote  [1]{``#1''}%
\providecommand \bibnamefont  [1]{#1}%
\providecommand \bibfnamefont [1]{#1}%
\providecommand \citenamefont [1]{#1}%
\providecommand \href@noop [0]{\@secondoftwo}%
\providecommand \href [0]{\begingroup \@sanitize@url \@href}%
\providecommand \@href[1]{\@@startlink{#1}\@@href}%
\providecommand \@@href[1]{\endgroup#1\@@endlink}%
\providecommand \@sanitize@url [0]{\catcode `\\12\catcode `\$12\catcode
  `\&12\catcode `\#12\catcode `\^12\catcode `\_12\catcode `\%12\relax}%
\providecommand \@@startlink[1]{}%
\providecommand \@@endlink[0]{}%
\providecommand \url  [0]{\begingroup\@sanitize@url \@url }%
\providecommand \@url [1]{\endgroup\@href {#1}{\urlprefix }}%
\providecommand \urlprefix  [0]{URL }%
\providecommand \Eprint [0]{\href }%
\providecommand \doibase [0]{http://dx.doi.org/}%
\providecommand \selectlanguage [0]{\@gobble}%
\providecommand \bibinfo  [0]{\@secondoftwo}%
\providecommand \bibfield  [0]{\@secondoftwo}%
\providecommand \translation [1]{[#1]}%
\providecommand \BibitemOpen [0]{}%
\providecommand \bibitemStop [0]{}%
\providecommand \bibitemNoStop [0]{.\EOS\space}%
\providecommand \EOS [0]{\spacefactor3000\relax}%
\providecommand \BibitemShut  [1]{\csname bibitem#1\endcsname}%
\let\auto@bib@innerbib\@empty
\bibitem [{\citenamefont {Feynman}(1982)}]{feynman82}%
  \BibitemOpen
  \bibfield  {author} {\bibinfo {author} {\bibfnamefont {R.~P.}\ \bibnamefont
  {Feynman}},\ }\href@noop {} {\bibfield  {journal} {\bibinfo  {journal} {Int.
  J. Theor. Phys.}\ }\textbf {\bibinfo {volume} {21}},\ \bibinfo {pages} {467}
  (\bibinfo {year} {1982})}\BibitemShut {NoStop}%
\bibitem [{\citenamefont {Georgescu}\ \emph {et~al.}(2014)\citenamefont
  {Georgescu}, \citenamefont {Ashhab},\ and\ \citenamefont
  {Nori}}]{georgescu2014quantum}%
  \BibitemOpen
  \bibfield  {author} {\bibinfo {author} {\bibfnamefont {I.~M.}\ \bibnamefont
  {Georgescu}}, \bibinfo {author} {\bibfnamefont {S.}~\bibnamefont {Ashhab}}, \
  and\ \bibinfo {author} {\bibfnamefont {F.}~\bibnamefont {Nori}},\ }\href@noop
  {} {\bibfield  {journal} {\bibinfo  {journal} {Rev. Mod. Phys.}\ }\textbf
  {\bibinfo {volume} {86}},\ \bibinfo {pages} {153} (\bibinfo {year}
  {2014})}\BibitemShut {NoStop}%
\bibitem [{\citenamefont {Aspuru-Guzik}\ and\ \citenamefont
  {Walther}(2012)}]{aspuru2012photonic}%
  \BibitemOpen
  \bibfield  {author} {\bibinfo {author} {\bibfnamefont {A.}~\bibnamefont
  {Aspuru-Guzik}}\ and\ \bibinfo {author} {\bibfnamefont {P.}~\bibnamefont
  {Walther}},\ }\href@noop {} {\bibfield  {journal} {\bibinfo  {journal} {Nat.
  Phys.}\ }\textbf {\bibinfo {volume} {8}},\ \bibinfo {pages} {285} (\bibinfo
  {year} {2012})}\BibitemShut {NoStop}%
\bibitem [{\citenamefont {Noh}\ and\ \citenamefont
  {Angelakis}(2016)}]{noh2016quantum}%
  \BibitemOpen
  \bibfield  {author} {\bibinfo {author} {\bibfnamefont {C.}~\bibnamefont
  {Noh}}\ and\ \bibinfo {author} {\bibfnamefont {D.~G.}\ \bibnamefont
  {Angelakis}},\ }\href@noop {} {\bibfield  {journal} {\bibinfo  {journal}
  {arXiv:1604.04433}\ } (\bibinfo {year} {2016})}\BibitemShut {NoStop}%
\bibitem [{\citenamefont {Heidemann}\ \emph {et~al.}(2007)\citenamefont
  {Heidemann}, \citenamefont {Raitzsch}, \citenamefont {Bendkowsky},
  \citenamefont {Butscher}, \citenamefont {L{\"o}w}, \citenamefont {Santos},\
  and\ \citenamefont {Pfau}}]{heidemann2007evidence}%
  \BibitemOpen
  \bibfield  {author} {\bibinfo {author} {\bibfnamefont {R.}~\bibnamefont
  {Heidemann}}, \bibinfo {author} {\bibfnamefont {U.}~\bibnamefont {Raitzsch}},
  \bibinfo {author} {\bibfnamefont {V.}~\bibnamefont {Bendkowsky}}, \bibinfo
  {author} {\bibfnamefont {B.}~\bibnamefont {Butscher}}, \bibinfo {author}
  {\bibfnamefont {R.}~\bibnamefont {L{\"o}w}}, \bibinfo {author} {\bibfnamefont
  {L.}~\bibnamefont {Santos}}, \ and\ \bibinfo {author} {\bibfnamefont
  {T.}~\bibnamefont {Pfau}},\ }\href@noop {} {\bibfield  {journal} {\bibinfo
  {journal} {Phys. Rev. Lett.}\ }\textbf {\bibinfo {volume} {99}},\ \bibinfo
  {pages} {163601} (\bibinfo {year} {2007})}\BibitemShut {NoStop}%
\bibitem [{\citenamefont {Peyronel}\ \emph {et~al.}(2012)\citenamefont
  {Peyronel}, \citenamefont {Firstenberg}, \citenamefont {Liang}, \citenamefont
  {Hofferberth}, \citenamefont {Gorshkov}, \citenamefont {Pohl}, \citenamefont
  {Lukin},\ and\ \citenamefont {Vuleti{\'c}}}]{peyronel2012quantum}%
  \BibitemOpen
  \bibfield  {author} {\bibinfo {author} {\bibfnamefont {T.}~\bibnamefont
  {Peyronel}}, \bibinfo {author} {\bibfnamefont {O.}~\bibnamefont
  {Firstenberg}}, \bibinfo {author} {\bibfnamefont {Q.-Y.}\ \bibnamefont
  {Liang}}, \bibinfo {author} {\bibfnamefont {S.}~\bibnamefont {Hofferberth}},
  \bibinfo {author} {\bibfnamefont {A.~V.}\ \bibnamefont {Gorshkov}}, \bibinfo
  {author} {\bibfnamefont {T.}~\bibnamefont {Pohl}}, \bibinfo {author}
  {\bibfnamefont {M.~D.}\ \bibnamefont {Lukin}}, \ and\ \bibinfo {author}
  {\bibfnamefont {V.}~\bibnamefont {Vuleti{\'c}}},\ }\href@noop {} {\bibfield
  {journal} {\bibinfo  {journal} {Nature (London)}\ }\textbf {\bibinfo {volume}
  {488}},\ \bibinfo {pages} {57} (\bibinfo {year} {2012})}\BibitemShut
  {NoStop}%
\bibitem [{\citenamefont {Houck}\ \emph {et~al.}(2012)\citenamefont {Houck},
  \citenamefont {T{\"u}reci},\ and\ \citenamefont {Koch}}]{houck2012chip}%
  \BibitemOpen
  \bibfield  {author} {\bibinfo {author} {\bibfnamefont {A.~A.}\ \bibnamefont
  {Houck}}, \bibinfo {author} {\bibfnamefont {H.~E.}\ \bibnamefont
  {T{\"u}reci}}, \ and\ \bibinfo {author} {\bibfnamefont {J.}~\bibnamefont
  {Koch}},\ }\href@noop {} {\bibfield  {journal} {\bibinfo  {journal} {Nat.
  Phys.}\ }\textbf {\bibinfo {volume} {8}},\ \bibinfo {pages} {292} (\bibinfo
  {year} {2012})}\BibitemShut {NoStop}%
\bibitem [{\citenamefont {Hennessy}\ \emph {et~al.}(2007)\citenamefont
  {Hennessy}, \citenamefont {Badolato}, \citenamefont {Winger}, \citenamefont
  {Gerace}, \citenamefont {Atat{\"{u}}re}, \citenamefont {Gulde}, \citenamefont
  {F{\"{a}}lt}, \citenamefont {Hu},\ and\ \citenamefont {Imamo{\u
  g}lu}}]{Hennessy2007}%
  \BibitemOpen
  \bibfield  {author} {\bibinfo {author} {\bibfnamefont {K.}~\bibnamefont
  {Hennessy}}, \bibinfo {author} {\bibfnamefont {A.}~\bibnamefont {Badolato}},
  \bibinfo {author} {\bibfnamefont {M.}~\bibnamefont {Winger}}, \bibinfo
  {author} {\bibfnamefont {D.}~\bibnamefont {Gerace}}, \bibinfo {author}
  {\bibfnamefont {M.}~\bibnamefont {Atat{\"{u}}re}}, \bibinfo {author}
  {\bibfnamefont {S.}~\bibnamefont {Gulde}}, \bibinfo {author} {\bibfnamefont
  {S.}~\bibnamefont {F{\"{a}}lt}}, \bibinfo {author} {\bibfnamefont {E.~L.}\
  \bibnamefont {Hu}}, \ and\ \bibinfo {author} {\bibfnamefont {A.}~\bibnamefont
  {Imamo{\u g}lu}},\ }\href {\doibase 10.1038/nature05586} {\bibfield
  {journal} {\bibinfo  {journal} {Nature (London)}\ }\textbf {\bibinfo {volume}
  {445}},\ \bibinfo {pages} {896} (\bibinfo {year} {2007})}\BibitemShut
  {NoStop}%
\bibitem [{\citenamefont {Kim}\ \emph {et~al.}(2013)\citenamefont {Kim},
  \citenamefont {Bose}, \citenamefont {Shen}, \citenamefont {Solomon},\ and\
  \citenamefont {Waks}}]{kim.2013.373--377}%
  \BibitemOpen
  \bibfield  {author} {\bibinfo {author} {\bibfnamefont {H.}~\bibnamefont
  {Kim}}, \bibinfo {author} {\bibfnamefont {R.}~\bibnamefont {Bose}}, \bibinfo
  {author} {\bibfnamefont {T.~C.}\ \bibnamefont {Shen}}, \bibinfo {author}
  {\bibfnamefont {G.~S.}\ \bibnamefont {Solomon}}, \ and\ \bibinfo {author}
  {\bibfnamefont {E.}~\bibnamefont {Waks}},\ }\href {\doibase
  10.1038/nphoton.2013.48} {\bibfield  {journal} {\bibinfo  {journal} {Nat.
  Photonics}\ }\textbf {\bibinfo {volume} {7}},\ \bibinfo {pages} {373}
  (\bibinfo {year} {2013})}\BibitemShut {NoStop}%
\bibitem [{\citenamefont {Chang}\ \emph {et~al.}(2014)\citenamefont {Chang},
  \citenamefont {Vuletic},\ and\ \citenamefont {Lukin}}]{RYDBERG}%
  \BibitemOpen
  \bibfield  {author} {\bibinfo {author} {\bibfnamefont {D.~E.}\ \bibnamefont
  {Chang}}, \bibinfo {author} {\bibfnamefont {V.}~\bibnamefont {Vuletic}}, \
  and\ \bibinfo {author} {\bibfnamefont {M.~D.}\ \bibnamefont {Lukin}},\ }\href
  {http://dx.doi.org/10.1038/nphoton.2014.192} {\bibfield  {journal} {\bibinfo
  {journal} {Nat. Photonics}\ }\textbf {\bibinfo {volume} {8}},\ \bibinfo
  {pages} {685} (\bibinfo {year} {2014})}\BibitemShut {NoStop}%
\bibitem [{\citenamefont {Kasprzak}\ \emph {et~al.}(2006)\citenamefont
  {Kasprzak}, \citenamefont {Richard}, \citenamefont {Kundermann},
  \citenamefont {Baas}, \citenamefont {Jeambrun}, \citenamefont {Keeling},
  \citenamefont {Marchetti}, \citenamefont {Szyma{\'n}ska}, \citenamefont
  {Andre}, \citenamefont {Staehli}, \citenamefont {Savona}, \citenamefont
  {Littlewood}, \citenamefont {Deveaud},\ and\ \citenamefont
  {Dang}}]{kasprzak2006bose}%
  \BibitemOpen
  \bibfield  {author} {\bibinfo {author} {\bibfnamefont {J.}~\bibnamefont
  {Kasprzak}}, \bibinfo {author} {\bibfnamefont {M.}~\bibnamefont {Richard}},
  \bibinfo {author} {\bibfnamefont {S.}~\bibnamefont {Kundermann}}, \bibinfo
  {author} {\bibfnamefont {A.}~\bibnamefont {Baas}}, \bibinfo {author}
  {\bibfnamefont {P.}~\bibnamefont {Jeambrun}}, \bibinfo {author}
  {\bibfnamefont {J.~M.~J.}\ \bibnamefont {Keeling}}, \bibinfo {author}
  {\bibfnamefont {F.~M.}\ \bibnamefont {Marchetti}}, \bibinfo {author}
  {\bibfnamefont {M.~H.}\ \bibnamefont {Szyma{\'n}ska}}, \bibinfo {author}
  {\bibfnamefont {R.}~\bibnamefont {Andre}}, \bibinfo {author} {\bibfnamefont
  {J.~L.}\ \bibnamefont {Staehli}}, \bibinfo {author} {\bibfnamefont
  {V.}~\bibnamefont {Savona}}, \bibinfo {author} {\bibfnamefont {P.~B.}\
  \bibnamefont {Littlewood}}, \bibinfo {author} {\bibfnamefont
  {B.}~\bibnamefont {Deveaud}}, \ and\ \bibinfo {author} {\bibfnamefont
  {L.~S.}\ \bibnamefont {Dang}},\ }\href@noop {} {\bibfield  {journal}
  {\bibinfo  {journal} {Nature (London)}\ }\textbf {\bibinfo {volume} {443}},\
  \bibinfo {pages} {409} (\bibinfo {year} {2006})}\BibitemShut {NoStop}%
\bibitem [{\citenamefont {Klaers}\ \emph {et~al.}(2010)\citenamefont {Klaers},
  \citenamefont {Schmitt}, \citenamefont {Vewinger},\ and\ \citenamefont
  {Weitz}}]{klaers2010bose}%
  \BibitemOpen
  \bibfield  {author} {\bibinfo {author} {\bibfnamefont {J.}~\bibnamefont
  {Klaers}}, \bibinfo {author} {\bibfnamefont {J.}~\bibnamefont {Schmitt}},
  \bibinfo {author} {\bibfnamefont {F.}~\bibnamefont {Vewinger}}, \ and\
  \bibinfo {author} {\bibfnamefont {M.}~\bibnamefont {Weitz}},\ }\href@noop {}
  {\bibfield  {journal} {\bibinfo  {journal} {Nature (London)}\ }\textbf
  {\bibinfo {volume} {468}},\ \bibinfo {pages} {545} (\bibinfo {year}
  {2010})}\BibitemShut {NoStop}%
\bibitem [{\citenamefont {Raghu}\ and\ \citenamefont
  {Haldane}(2008)}]{Raghu2008}%
  \BibitemOpen
  \bibfield  {author} {\bibinfo {author} {\bibfnamefont {S.}~\bibnamefont
  {Raghu}}\ and\ \bibinfo {author} {\bibfnamefont {F.}~\bibnamefont
  {Haldane}},\ }\href {\doibase 10.1103/PhysRevA.78.033834} {\bibfield
  {journal} {\bibinfo  {journal} {Phys. Rev. A}\ }\textbf {\bibinfo {volume}
  {78}},\ \bibinfo {pages} {033834} (\bibinfo {year} {2008})}\BibitemShut
  {NoStop}%
\bibitem [{\citenamefont {Hafezi}\ \emph {et~al.}(2011)\citenamefont {Hafezi},
  \citenamefont {Demler}, \citenamefont {Lukin},\ and\ \citenamefont
  {Taylor}}]{Hafezi2011}%
  \BibitemOpen
  \bibfield  {author} {\bibinfo {author} {\bibfnamefont {M.}~\bibnamefont
  {Hafezi}}, \bibinfo {author} {\bibfnamefont {E.~A.}\ \bibnamefont {Demler}},
  \bibinfo {author} {\bibfnamefont {M.~D.}\ \bibnamefont {Lukin}}, \ and\
  \bibinfo {author} {\bibfnamefont {J.~M.}\ \bibnamefont {Taylor}},\ }\href
  {\doibase 10.1038/nphys2063} {\bibfield  {journal} {\bibinfo  {journal} {Nat.
  Phys.}\ }\textbf {\bibinfo {volume} {7}},\ \bibinfo {pages} {907} (\bibinfo
  {year} {2011})}\BibitemShut {NoStop}%
\bibitem [{\citenamefont {Fang}\ \emph {et~al.}(2012)\citenamefont {Fang},
  \citenamefont {Yu},\ and\ \citenamefont {Fan}}]{Fang2012}%
  \BibitemOpen
  \bibfield  {author} {\bibinfo {author} {\bibfnamefont {K.}~\bibnamefont
  {Fang}}, \bibinfo {author} {\bibfnamefont {Z.}~\bibnamefont {Yu}}, \ and\
  \bibinfo {author} {\bibfnamefont {S.}~\bibnamefont {Fan}},\ }\href {\doibase
  10.1038/NPHOTON.2012.236} {\bibfield  {journal} {\bibinfo  {journal} {Nat.
  Photon.}\ }\textbf {\bibinfo {volume} {6}},\ \bibinfo {pages} {782} (\bibinfo
  {year} {2012})}\BibitemShut {NoStop}%
\bibitem [{\citenamefont {Balili}\ \emph {et~al.}(2007)\citenamefont {Balili},
  \citenamefont {Hartwell}, \citenamefont {Snoke}, \citenamefont {Pfeiffer},\
  and\ \citenamefont {West}}]{balili2007bose}%
  \BibitemOpen
  \bibfield  {author} {\bibinfo {author} {\bibfnamefont {R.}~\bibnamefont
  {Balili}}, \bibinfo {author} {\bibfnamefont {V.}~\bibnamefont {Hartwell}},
  \bibinfo {author} {\bibfnamefont {D.}~\bibnamefont {Snoke}}, \bibinfo
  {author} {\bibfnamefont {L.}~\bibnamefont {Pfeiffer}}, \ and\ \bibinfo
  {author} {\bibfnamefont {K.}~\bibnamefont {West}},\ }\href@noop {} {\bibfield
   {journal} {\bibinfo  {journal} {Science}\ }\textbf {\bibinfo {volume}
  {316}},\ \bibinfo {pages} {1007} (\bibinfo {year} {2007})}\BibitemShut
  {NoStop}%
\bibitem [{\citenamefont {Deng}\ \emph {et~al.}(2010)\citenamefont {Deng},
  \citenamefont {Haug},\ and\ \citenamefont {Yamamoto}}]{deng2010exciton}%
  \BibitemOpen
  \bibfield  {author} {\bibinfo {author} {\bibfnamefont {H.}~\bibnamefont
  {Deng}}, \bibinfo {author} {\bibfnamefont {H.}~\bibnamefont {Haug}}, \ and\
  \bibinfo {author} {\bibfnamefont {Y.}~\bibnamefont {Yamamoto}},\ }\href@noop
  {} {\bibfield  {journal} {\bibinfo  {journal} {Rev. Mod. Phys.}\ }\textbf
  {\bibinfo {volume} {82}},\ \bibinfo {pages} {1489} (\bibinfo {year}
  {2010})}\BibitemShut {NoStop}%
\bibitem [{\citenamefont {Tosi}\ \emph {et~al.}(2012)\citenamefont {Tosi},
  \citenamefont {Christmann}, \citenamefont {Berloff}, \citenamefont {Tsotsis},
  \citenamefont {Gao}, \citenamefont {Hatzopoulos}, \citenamefont {Savvidis},\
  and\ \citenamefont {Baumberg}}]{tosi.2012.190--194}%
  \BibitemOpen
  \bibfield  {author} {\bibinfo {author} {\bibfnamefont {G.}~\bibnamefont
  {Tosi}}, \bibinfo {author} {\bibfnamefont {G.}~\bibnamefont {Christmann}},
  \bibinfo {author} {\bibfnamefont {N.~G.}\ \bibnamefont {Berloff}}, \bibinfo
  {author} {\bibfnamefont {P.}~\bibnamefont {Tsotsis}}, \bibinfo {author}
  {\bibfnamefont {T.}~\bibnamefont {Gao}}, \bibinfo {author} {\bibfnamefont
  {Z.}~\bibnamefont {Hatzopoulos}}, \bibinfo {author} {\bibfnamefont {P.~G.}\
  \bibnamefont {Savvidis}}, \ and\ \bibinfo {author} {\bibfnamefont {J.~J.}\
  \bibnamefont {Baumberg}},\ }\href {\doibase 10.1038/nphys2182} {\bibfield
  {journal} {\bibinfo  {journal} {Nat. Phys.}\ }\textbf {\bibinfo {volume}
  {8}},\ \bibinfo {pages} {190} (\bibinfo {year} {2012})}\BibitemShut {NoStop}%
\bibitem [{\citenamefont {Keeling}\ and\ \citenamefont
  {Berloff}(2011)}]{keeling.2011.131--151}%
  \BibitemOpen
  \bibfield  {author} {\bibinfo {author} {\bibfnamefont {J.}~\bibnamefont
  {Keeling}}\ and\ \bibinfo {author} {\bibfnamefont {N.~G.}\ \bibnamefont
  {Berloff}},\ }\href {\doibase 10.1080/00107514.2010.550120} {\bibfield
  {journal} {\bibinfo  {journal} {Contemp. Phys.}\ }\textbf {\bibinfo {volume}
  {52}},\ \bibinfo {pages} {131} (\bibinfo {year} {2011})}\BibitemShut
  {NoStop}%
\bibitem [{\citenamefont {Nissen}\ \emph {et~al.}(2012)\citenamefont {Nissen},
  \citenamefont {Schmidt}, \citenamefont {Biondi}, \citenamefont {Blatter},
  \citenamefont {T{\"{u}}reci},\ and\ \citenamefont {Keeling}}]{nissen.2012.}%
  \BibitemOpen
  \bibfield  {author} {\bibinfo {author} {\bibfnamefont {F.}~\bibnamefont
  {Nissen}}, \bibinfo {author} {\bibfnamefont {S.}~\bibnamefont {Schmidt}},
  \bibinfo {author} {\bibfnamefont {M.}~\bibnamefont {Biondi}}, \bibinfo
  {author} {\bibfnamefont {G.}~\bibnamefont {Blatter}}, \bibinfo {author}
  {\bibfnamefont {H.~E.}\ \bibnamefont {T{\"{u}}reci}}, \ and\ \bibinfo
  {author} {\bibfnamefont {J.}~\bibnamefont {Keeling}},\ }\href
  {http://arxiv.org/abs/1202.1961} {\bibfield  {journal} {\bibinfo  {journal}
  {Phys. Rev. Lett.}\ }\textbf {\bibinfo {volume} {108}},\ \bibinfo {pages}
  {233603} (\bibinfo {year} {2012})}\BibitemShut {NoStop}%
\bibitem [{\citenamefont {Landauer}(1957)}]{landauer1957spatial}%
  \BibitemOpen
  \bibfield  {author} {\bibinfo {author} {\bibfnamefont {R.}~\bibnamefont
  {Landauer}},\ }\href@noop {} {\bibfield  {journal} {\bibinfo  {journal} {IBM
  J. Res. Dev.}\ }\textbf {\bibinfo {volume} {1}},\ \bibinfo {pages} {223}
  (\bibinfo {year} {1957})}\BibitemShut {NoStop}%
\bibitem [{\citenamefont {Landauer}(1970)}]{landauer1970electrical}%
  \BibitemOpen
  \bibfield  {author} {\bibinfo {author} {\bibfnamefont {R.}~\bibnamefont
  {Landauer}},\ }\href@noop {} {\bibfield  {journal} {\bibinfo  {journal}
  {Philos. Mag.}\ }\textbf {\bibinfo {volume} {21}},\ \bibinfo {pages} {863}
  (\bibinfo {year} {1970})}\BibitemShut {NoStop}%
\bibitem [{\citenamefont {Imry}\ and\ \citenamefont
  {Landauer}(1999)}]{imry1999conductance}%
  \BibitemOpen
  \bibfield  {author} {\bibinfo {author} {\bibfnamefont {Y.}~\bibnamefont
  {Imry}}\ and\ \bibinfo {author} {\bibfnamefont {R.}~\bibnamefont
  {Landauer}},\ }\href@noop {} {\bibfield  {journal} {\bibinfo  {journal} {Rev.
  Mod. Phys.}\ }\textbf {\bibinfo {volume} {71}},\ \bibinfo {pages} {S306}
  (\bibinfo {year} {1999})}\BibitemShut {NoStop}%
\bibitem [{\citenamefont {B{\"u}ttiker}(1986)}]{buttiker1986four}%
  \BibitemOpen
  \bibfield  {author} {\bibinfo {author} {\bibfnamefont {M.}~\bibnamefont
  {B{\"u}ttiker}},\ }\href@noop {} {\bibfield  {journal} {\bibinfo  {journal}
  {Phys. Rev. Lett.}\ }\textbf {\bibinfo {volume} {57}},\ \bibinfo {pages}
  {1761} (\bibinfo {year} {1986})}\BibitemShut {NoStop}%
\bibitem [{\citenamefont {Caroli}\ \emph {et~al.}(1971)\citenamefont {Caroli},
  \citenamefont {Combescot}, \citenamefont {Nozieres},\ and\ \citenamefont
  {Saint-James}}]{caroli1971direct}%
  \BibitemOpen
  \bibfield  {author} {\bibinfo {author} {\bibfnamefont {C.}~\bibnamefont
  {Caroli}}, \bibinfo {author} {\bibfnamefont {R.}~\bibnamefont {Combescot}},
  \bibinfo {author} {\bibfnamefont {P.}~\bibnamefont {Nozieres}}, \ and\
  \bibinfo {author} {\bibfnamefont {D.}~\bibnamefont {Saint-James}},\
  }\href@noop {} {\bibfield  {journal} {\bibinfo  {journal} {J. Phys. C}\
  }\textbf {\bibinfo {volume} {4}},\ \bibinfo {pages} {916} (\bibinfo {year}
  {1971})}\BibitemShut {NoStop}%
\bibitem [{\citenamefont {Meir}\ and\ \citenamefont
  {Wingreen}(1992)}]{meir1992landauer}%
  \BibitemOpen
  \bibfield  {author} {\bibinfo {author} {\bibfnamefont {Y.}~\bibnamefont
  {Meir}}\ and\ \bibinfo {author} {\bibfnamefont {N.~S.}\ \bibnamefont
  {Wingreen}},\ }\href@noop {} {\bibfield  {journal} {\bibinfo  {journal}
  {Phys. Rev. Lett.}\ }\textbf {\bibinfo {volume} {68}},\ \bibinfo {pages}
  {2512} (\bibinfo {year} {1992})}\BibitemShut {NoStop}%
\bibitem [{\citenamefont {Haug}\ \emph {et~al.}(2008)\citenamefont {Haug},
  \citenamefont {Jauho},\ and\ \citenamefont {Cardona}}]{haug2008quantum}%
  \BibitemOpen
  \bibfield  {author} {\bibinfo {author} {\bibfnamefont {H.}~\bibnamefont
  {Haug}}, \bibinfo {author} {\bibfnamefont {A.-P.}\ \bibnamefont {Jauho}}, \
  and\ \bibinfo {author} {\bibfnamefont {M.}~\bibnamefont {Cardona}},\
  }\href@noop {} {\emph {\bibinfo {title} {Quantum Kinetics in Transport and
  Optics of Semiconductors}}},\ Vol.~\bibinfo {volume} {14}\ (\bibinfo
  {publisher} {Springer, Berlin Heidelberg},\ \bibinfo {year}
  {2008})\BibitemShut {NoStop}%
\bibitem [{\citenamefont {Keldysh}(1965)}]{keldysh1965diagram}%
  \BibitemOpen
  \bibfield  {author} {\bibinfo {author} {\bibfnamefont {L.~V.}\ \bibnamefont
  {Keldysh}},\ }\href@noop {} {\bibfield  {journal} {\bibinfo  {journal} {Sov.
  Phys. JETP}\ }\textbf {\bibinfo {volume} {20}},\ \bibinfo {pages} {1018}
  (\bibinfo {year} {1965})}\BibitemShut {NoStop}%
\bibitem [{\citenamefont {Langreth}(1976)}]{langreth1976linear}%
  \BibitemOpen
  \bibfield  {author} {\bibinfo {author} {\bibfnamefont {D.~C.}\ \bibnamefont
  {Langreth}},\ }in\ \href@noop {} {\emph {\bibinfo {booktitle} {Linear and
  Nonlinear Electron Transport in Solids}}},\ \bibinfo {editor} {edited by\
  \bibinfo {editor} {\bibfnamefont {J.~T.}\ \bibnamefont {Devreese}}\ and\
  \bibinfo {editor} {\bibfnamefont {V.~E.}\ \bibnamefont {van Doren}}}\
  (\bibinfo  {publisher} {Plenum, New York},\ \bibinfo {year} {1976})\ pp.\
  \bibinfo {pages} {3--32}\BibitemShut {NoStop}%
\bibitem [{\citenamefont {Wurfel}(1982)}]{wurfel1982chemical}%
  \BibitemOpen
  \bibfield  {author} {\bibinfo {author} {\bibfnamefont {P.}~\bibnamefont
  {Wurfel}},\ }\href@noop {} {\bibfield  {journal} {\bibinfo  {journal} {J.
  Phys. C}\ }\textbf {\bibinfo {volume} {15}},\ \bibinfo {pages} {3967}
  (\bibinfo {year} {1982})}\BibitemShut {NoStop}%
\bibitem [{\citenamefont {Herrmann}\ and\ \citenamefont
  {W{\"u}rfel}(2005)}]{herrmann2005light}%
  \BibitemOpen
  \bibfield  {author} {\bibinfo {author} {\bibfnamefont {F.}~\bibnamefont
  {Herrmann}}\ and\ \bibinfo {author} {\bibfnamefont {P.}~\bibnamefont
  {W{\"u}rfel}},\ }\href@noop {} {\bibfield  {journal} {\bibinfo  {journal}
  {Am. J. Phys.}\ }\textbf {\bibinfo {volume} {73}},\ \bibinfo {pages} {717}
  (\bibinfo {year} {2005})}\BibitemShut {NoStop}%
\bibitem [{\citenamefont {Schmitt}\ \emph {et~al.}(2014)\citenamefont
  {Schmitt}, \citenamefont {Damm}, \citenamefont {Dung}, \citenamefont
  {Vewinger}, \citenamefont {Klaers},\ and\ \citenamefont
  {Weitz}}]{schmitt2014observation}%
  \BibitemOpen
  \bibfield  {author} {\bibinfo {author} {\bibfnamefont {J.}~\bibnamefont
  {Schmitt}}, \bibinfo {author} {\bibfnamefont {T.}~\bibnamefont {Damm}},
  \bibinfo {author} {\bibfnamefont {D.}~\bibnamefont {Dung}}, \bibinfo {author}
  {\bibfnamefont {F.}~\bibnamefont {Vewinger}}, \bibinfo {author}
  {\bibfnamefont {J.}~\bibnamefont {Klaers}}, \ and\ \bibinfo {author}
  {\bibfnamefont {M.}~\bibnamefont {Weitz}},\ }\href {\doibase
  10.1103/PhysRevLett.112.030401} {\bibfield  {journal} {\bibinfo  {journal}
  {Phys. Rev. Lett.}\ }\textbf {\bibinfo {volume} {112}},\ \bibinfo {pages}
  {030401} (\bibinfo {year} {2014})}\BibitemShut {NoStop}%
\bibitem [{\citenamefont {Ries}\ and\ \citenamefont
  {McEvoy}(1991)}]{ries1991chemical}%
  \BibitemOpen
  \bibfield  {author} {\bibinfo {author} {\bibfnamefont {H.}~\bibnamefont
  {Ries}}\ and\ \bibinfo {author} {\bibfnamefont {A.~J.}\ \bibnamefont
  {McEvoy}},\ }\href@noop {} {\bibfield  {journal} {\bibinfo  {journal} {J.
  Photochem. Photobiol. Chem.}\ }\textbf {\bibinfo {volume} {59}},\ \bibinfo
  {pages} {11} (\bibinfo {year} {1991})}\BibitemShut {NoStop}%
\bibitem [{\citenamefont {Smestad}\ and\ \citenamefont
  {Ries}(1992)}]{smestad1992luminescence}%
  \BibitemOpen
  \bibfield  {author} {\bibinfo {author} {\bibfnamefont {G.}~\bibnamefont
  {Smestad}}\ and\ \bibinfo {author} {\bibfnamefont {H.}~\bibnamefont {Ries}},\
  }\href@noop {} {\bibfield  {journal} {\bibinfo  {journal} {Sol. Energy Mater.
  Sol. Cells}\ }\textbf {\bibinfo {volume} {25}},\ \bibinfo {pages} {51}
  (\bibinfo {year} {1992})}\BibitemShut {NoStop}%
\bibitem [{\citenamefont {Berdahl}(1985)}]{berdahl1985radiant}%
  \BibitemOpen
  \bibfield  {author} {\bibinfo {author} {\bibfnamefont {P.}~\bibnamefont
  {Berdahl}},\ }\href@noop {} {\bibfield  {journal} {\bibinfo  {journal} {J.
  Appl. Phys.}\ }\textbf {\bibinfo {volume} {58}},\ \bibinfo {pages} {1369}
  (\bibinfo {year} {1985})}\BibitemShut {NoStop}%
\bibitem [{\citenamefont {Hafezi}\ \emph {et~al.}(2015)\citenamefont {Hafezi},
  \citenamefont {Adhikari},\ and\ \citenamefont {Taylor}}]{hafezi2015chemical}%
  \BibitemOpen
  \bibfield  {author} {\bibinfo {author} {\bibfnamefont {M.}~\bibnamefont
  {Hafezi}}, \bibinfo {author} {\bibfnamefont {P.}~\bibnamefont {Adhikari}}, \
  and\ \bibinfo {author} {\bibfnamefont {J.~M.}\ \bibnamefont {Taylor}},\
  }\href@noop {} {\bibfield  {journal} {\bibinfo  {journal} {Phys. Rev. B}\
  }\textbf {\bibinfo {volume} {92}},\ \bibinfo {pages} {174305} (\bibinfo
  {year} {2015})}\BibitemShut {NoStop}%
\bibitem [{\citenamefont {Wang}\ \emph {et~al.}(2014)\citenamefont {Wang},
  \citenamefont {Agarwalla}, \citenamefont {Li},\ and\ \citenamefont
  {Thingna}}]{wang2014nonequilibrium}%
  \BibitemOpen
  \bibfield  {author} {\bibinfo {author} {\bibfnamefont {J.-S.}\ \bibnamefont
  {Wang}}, \bibinfo {author} {\bibfnamefont {B.~K.}\ \bibnamefont {Agarwalla}},
  \bibinfo {author} {\bibfnamefont {H.}~\bibnamefont {Li}}, \ and\ \bibinfo
  {author} {\bibfnamefont {J.}~\bibnamefont {Thingna}},\ }\href@noop {}
  {\bibfield  {journal} {\bibinfo  {journal} {Front. Phys.}\ }\textbf {\bibinfo
  {volume} {9}},\ \bibinfo {pages} {673} (\bibinfo {year} {2014})}\BibitemShut
  {NoStop}%
\bibitem [{\citenamefont {Milburn}(1984)}]{milburn1984multimode}%
  \BibitemOpen
  \bibfield  {author} {\bibinfo {author} {\bibfnamefont {G.}~\bibnamefont
  {Milburn}},\ }\href@noop {} {\bibfield  {journal} {\bibinfo  {journal} {J.
  Phys. A}\ }\textbf {\bibinfo {volume} {17}},\ \bibinfo {pages} {737}
  (\bibinfo {year} {1984})}\BibitemShut {NoStop}%
\bibitem [{\citenamefont {Lee}\ and\ \citenamefont
  {Ramakrishnan}(1985)}]{lee1985disordered}%
  \BibitemOpen
  \bibfield  {author} {\bibinfo {author} {\bibfnamefont {P.~A.}\ \bibnamefont
  {Lee}}\ and\ \bibinfo {author} {\bibfnamefont {T.}~\bibnamefont
  {Ramakrishnan}},\ }\href@noop {} {\bibfield  {journal} {\bibinfo  {journal}
  {Rev. Mod. Phys.}\ }\textbf {\bibinfo {volume} {57}},\ \bibinfo {pages} {287}
  (\bibinfo {year} {1985})}\BibitemShut {NoStop}%
\bibitem [{\citenamefont {Imry}(2002)}]{imry2002introduction}%
  \BibitemOpen
  \bibfield  {author} {\bibinfo {author} {\bibfnamefont {Y.}~\bibnamefont
  {Imry}},\ }\href@noop {} {\emph {\bibinfo {title} {Introduction to Mesoscopic
  Physics}}},\ \bibinfo {edition} {2nd}\ ed.\ (\bibinfo  {publisher} {Oxford
  University Press, Oxford, UK},\ \bibinfo {year} {2002})\BibitemShut {NoStop}%
\bibitem [{\citenamefont {Akkermans}\ and\ \citenamefont
  {Montambaux}(2007)}]{akkermans2007mesoscopic}%
  \BibitemOpen
  \bibfield  {author} {\bibinfo {author} {\bibfnamefont {E.}~\bibnamefont
  {Akkermans}}\ and\ \bibinfo {author} {\bibfnamefont {G.}~\bibnamefont
  {Montambaux}},\ }\href@noop {} {\emph {\bibinfo {title} {Mesoscopic Physics
  of Electrons and Photons}}}\ (\bibinfo  {publisher} {Cambridge University
  Press, Cambridge, UK},\ \bibinfo {year} {2007})\BibitemShut {NoStop}%
\end{thebibliography}%

\end{document}